\documentclass[prb,preprint]{revtex4}

\pdfoutput=1

\usepackage{graphicx}

\begin{document}


\title{Determination of doped charge density in superconducting cuprates from NMR or stripes}
\medskip 

\date{December 3, 2020} \bigskip

\author{Manfred Bucher \\}
\affiliation{\text{\textnormal{Physics Department, California State University,}} \textnormal{Fresno,}
\textnormal{Fresno, California 93740-8031} \\}

\begin{abstract}
Independent investigations of nuclear quadrupole resonance (NQR) and of stripes in high-$T_c$ cuprates find a small deviation of doped-hole density $h$ from the doping level of $La_{2-x}Sr_xCuO_4$. The value observed with NQR, $ x - h \approx 0.02$, agrees closely with the density of itinerant holes, $\tilde{p}$, responsible for suppression of 3D-AFM, as obtained from stripe incommensurability.
The stripe model's assumption that doped holes in $La_{2-x}Sr_xCuO_4$ reside at oxygen sites, and that doped electrons in  $Ln_{2-x}Ce_xCuO_4$ ($Ln = Pr, Nd$) reside at copper sites, is (to a large degree) confirmed with NQR. The NQR finding of doped-hole probabilities in oxygen and copper   orbitals of $HgBa_2CuO_{4+\delta}$ and other oxygen-enriched high-$T_c$ cuprates, $P_p \simeq P_d \simeq 1/2$, as well as of oxygen-doped $YBa_2Cu_3O_{6+y}$, $P_p \simeq 2P_d \simeq 2/3$, is interpreted with the stripe model in terms of excess oxygen atoms in the $CuO_2$ planes and $CuO$ chains.
\end{abstract}
\maketitle

\pagebreak

\section{INTRODUCTION}
Superconductivity in doped copper oxides of high transition temperature (high $T_c$) occurs in the $CuO_2$ planes within certain doping ranges.
When doped with heterovalent metal, such as hole-doped $La_{2-x}Sr_xCuO_4$ or electron-doped $Pr_{2-x}Ce_xCuO_4$, the doped-charge density in the $CuO_2$ planes is readily given by stoichiometry, $p=x$ and $n=x$, respectively.
For oxygen-doped $YBa_2Cu_3O_{6+y}$, on the other hand, $p(y)$ cannot be inferred from stoichiometry because of simultaneous filling of $CuO$ chains. Instead, one needs to resort to indirect methods.
Most commonly used is the the ``universal-dome method'' which
obtains $p(y)$ from experimental data of $T_c(y)$ and $T_{c,max}$.\cite{1} Because of its crucial role in superconductivity, independent methods to determine the doped-charge density are valuable. Two such methods are discussed and compared here---one using nuclear quadrupole resonance, the other one stripe incommensurability.

Nuclear quadrupole resonance (NQR) arises from the interaction of the quadrupole moment of nuclear charge distribution  with the electric field gradient (EFG) at the nucleus. 
For nuclei of nonvanishing quadrupole moments, which holds for nuclear quantum numbers $I > \frac{1}{2}$, the method can provide information about the charge distribution in a compound. Particularly, with  $I(^{43}Cu)=\frac{3}{2}$ and $I(^{17}O)=\frac{5}{2}$, NQR can give the doped-charge density in the $CuO_2$ planes. This was shown in a seminal paper by Haase \textit{et al.} for $La_{2-x}Sr_xCuO_4$ and $YBa_2Cu_3O_{6+y}$ a decade and a half ago,\cite{2} followed by the same treatment of oxygen-enriched cuprates and of electron-doped $Ln_{2-x}Ce_xCuO_{4+\delta}$ ($Ln = Pr, Nd$) a decade later.\cite{3,4,5}

The doped-hole density $p$ (and doped-electron density $n$) of high-$T_c$ cuprates can also be determined from the incommensurability of charge-order and magnetization stripes, as shown in two recent papers.\cite{6,7} 
Shedding new light on the NQR results, here we point out commonalities with the stripe analysis, but also aspects that have been overlooked in the earlier studies or went unexplained.

\section{NQR DETERMINATION OF DOPED-HOLE DENSITY IN $\mathbf{La_{2-x}Sr_xCuO_4}$ }

Haase \textit{et al.}\cite{2} posit the NQR frequency at the oxygen nucleus, $\nu_O$,  to be proportional to the hole density $n_p$ of the $O\; 2p$ orbital, 
\begin{equation}
\nu_O =  Q_O \times n_p + C_O\;.
\end{equation}
The coefficient $Q_O$ is obtained from quantum-mechanical expressions of electric hyperfine interaction for isolated oxygen ions. Based on electron hopping parameters, $n_p$ is determined for the undoped $CuO_2$ plane as $n_{p0} = 0.11$. Together with the experimental value of $\nu_O$ in the parent crystal, this enables the determination of the material-specific constant $C_O$. 
The NQR frequency of the $Cu$ nucleus depends linearly on the hole density $n_d$ of the $Cu\;3d_{x^2-y^2}$ orbital, but also, by a cross-term, on neighboring $O\;2p$ orbitals,
\begin{equation}
    \nu_{Cu} =  Q_{Cu} \times n_d -Q_{Cu}^O \times (8-4n_p) + C_{Cu}\;. 
\end{equation}
The coefficients $Q_{Cu}$ and $Q_{Cu}^O$ are obtained from quantum-mechanical expressions for isolated ions.
With $n_{d0} = 1 - 2n_{p0} = 0.78$ for the undoped crystals, the constant $C_{Cu}$ can be determined.
For the \textit{doped} crystals, Haase \textit{et al.} make the assumption
\begin{equation}
    h = n_d + 2n_p -1 \;.
\end{equation}
(In Ref. 2 the doped-hole density $h$ is denoted as $\delta$.) 
With experimental NQR frequencies  $\nu_{Cu}$ and $\nu_O$ of $La_{2-x}Sr_xCuO_4$, along with the calculated coefficients and constants, the quantities $n_d$, $n_p$ and $h$ can be determined. They are listed in Table I.
The doped-hole density $h$ is close to the $Sr$-doping, $h \approx x$,
which attests to the predictive power of the NQR approach. A closer look shows that $h$ is systematically slightly lower than $x$,
\begin{equation}
    x - h = \Delta h \simeq 0.02 \;.
\end{equation}
Is this an inaccuracy caused by the approximations involved or the result of some underlying physics? We shall return to this question.

\begin{table}[ht!]
\begin{tabular}{|p{1.15cm}|p{1.8cm}|p{1.6cm}|p{1.5cm}|p{1.5cm}|p{1.5cm}|p{1.5cm}|p{1.5cm}|}
 \hline  \hline
$\;\;\;x$&$\nu_{Cu}$ [MHz]&$\nu_O$ [MHz]&$\;\;\;n_d$&$\;\;\;n_p$&$\;\;\;\;h$&$\;x-h$&$\;\;\;\tilde{p}$\\
 \hline  \hline
$\;0.00$&$\;\;\;\;\;33.2$&$\;\;\;0.147$&$\;\;\;0.780$&$\;\;\;0.110$&$\;\;0.00$&$\;\;0.00$&\\  
$\;0.075$&$\;\;\;\;\;34.2$&$\;\;\;0.18$&$\;\;\;0.784$&$\;\;\;0.137$&$\;\;0.058$ &$\;\;0.017$&$\;\;0.020$  \\
$\;0.10$&$\;\;\;\;\;34.6$&$\;\;\;0.195$&$\;\;\;0.785$&$\;\;\;0.149$&$\;\;0.084$&$\;\;0.016$&$\;\;0.015$ \\
$\;0.15$&$\;\;\;\;\;35.8$&$\;\;\;0.215$&$\;\;\;0.794$&$\;\;\;0.166$&$\;\;0.125$&$\;\;0.025$&$\;\;0.015$ \\
$\;0.20$&$\;\;\;\;\;36.6$&$\;\;\;0.245$&$\;\;\;0.797$&$\;\;\;0.190$&$\;\;0.190$&$\;\;0.023$&$\;\;0.015$ \\
$\;0.24$&$\;\;\;\;\;37.4$&$\;\;\;0.28$&$\;\;\;0.798$&$\;\;\;0.219$&$\;\;0.236$&$\;\;0.004$&$\;\;0.015$ \\
 \hline   \hline
\end{tabular}
\caption{NQR frequencies $\nu_{Cu}$ and $\nu_O$, hole density $n_d$ and $n_p$ in $Cu\; 3d_{x^2-y^2}$ and $O\; 2p_c$ orbitals, respectively, and density $h$ of doped holes in the $CuO_2$ plane of $La_{2-x}Sr_xCuO_4$, obtained from NQR data (Ref. 2); difference from nominal $Sr$ doping, $x-h$; and density of itinerant holes, $\tilde{p}$, from stripe incommensurability, Ref. 6.}

\label{table:1}  \end{table}

The hole density in the  $Cu\;3d_{x^2-y^2}$ orbital arises from the contribution of the parent crystal and a doping-dependent term,
\begin{equation}
    n_d = n_{d0} + P_dh \;.
\end{equation}
Here $P_d$ is the probability of a doped hole to reside at the $Cu^{2+}$ ion. The corresponding expression for oxygen is
\begin{equation}
    n_p = n_{p0} + \frac{1}{2}P_ph \;,
\end{equation}
where the factor 1/2 accounts for the \textit{two} oxygen ions in the $CuO_2$ plane of the unit cell. From the data in Table I, $P_d \simeq 0.07$ and $P_p \simeq 0.93$ is obtained.\cite{2} This shows that in $La_{2-x}Sr_xCuO_4$ the doped holes in the $CuO_2$ plane reside almost entirely at the oxygen sites---a conclusion that still holds in view of a small uncertainty of $n_p$, as the constant $C_O$ in Eq. (1) can be determined only within bounds by the formalism of the NQR approach.\cite{4}

\section{STRIPES IN $\mathbf{La_{2-x}Sr_xCuO_4}$}

The unit cell of \emph{pristine} ${La_2CuO_4}$ has a central $CuO_2$ plane, sandwiched by $LaO$ planes. Consider stepwise ionization, where brackets indicate electron localization at atoms, both within the planes and by transfer from the $LaO$ planes to the $CuO_2$ plane:
\bigskip 
\newline
\noindent $LaO \;\;:\; La^{3+} + 3e^- + \;\;O \rightarrow  La^{3+} + [2e^- + O]\; + 
\downarrow \overline {e^-\;} | \rightarrow  La^{3+} + O^{2-}$
\newline
$CuO_2 :\; Cu^{2+} + 2e^- + 2O \rightarrow  Cu^{2+} + [2e^- + O] + \;\;\;O \;\;\;\rightarrow  O^{2-} \;+  Cu^{2+} + O^{2-}$
\newline
$LaO \;\;:\; La^{3+} + 3e^- + \;\;O \rightarrow  La^{3+} + [2e^- + O]\; + 
\uparrow \underline {e^-\;} | \rightarrow  La^{3+} + O^{2-}$
\bigskip

\noindent In the simplest case of {\it doping},
$Sr$ substitutes,  in some cells, $La$ in both sandwiching planes: 

\bigskip 
\noindent $SrO \;\;:\; Sr^{2+} + 2e^- + \;\;O \rightarrow  Sr^{2+} + [2e^- + O]\; \;\;\;\;\;\;\;\;\;\;\;\; \;\rightarrow  Sr^{2+} + O^{2-}$
\newline
$CuO_2 :\; Cu^{2+} + 2e^- + 2O \rightarrow  Cu^{2+} + [2e^- + O] + \;\;\;O \;\;\;\rightarrow  O^{2-} \;+  Cu^{2+} + \mathbf{\tilde{O}}$
\newline
$SrO \;\;:\; Sr^{2+} + 2e^- + \;\;O \rightarrow  Sr^{2+} + [2e^- + O]\; \;\;\;\;\;\;\;\;\;\;\;\;\; \rightarrow  Sr^{2+} + O^{2-}$
\bigskip

\noindent The lack of electron transfer from the sandwiching planes to the $CuO_2$ plane leaves some oxygen atoms \textit{neutral} (marked bold above and below). Compared to the host crystal, they can be regarded as housing the holes (pairwise).\cite{6} Up to a doping density $\tilde{p} \le  0.02$, such holes are \textit{itinerant}, enabling $\mathbf{\tilde{O}}$ atoms to skirmish long-range antiferromagnetism (3D-AFM).\cite{6} The remaining lack of electron transfer leaves more oxygen atoms \textit{stationary} at lattice sites, $\mathbf{O}$. They give rise to static stripes.

\bigskip 
\noindent $SrO \;\;:\; Sr^{2+} + 2e^- + \;\;O \rightarrow  Sr^{2+} + [2e^- + O]\; \;\;\;\;\;\;\;\;\;\;\;\; \;\rightarrow  Sr^{2+} + O^{2-}$
\newline
$CuO_2 :\; Cu^{2+} + 2e^- + 2O \rightarrow  Cu^{2+} + [2e^- + O] + \;\;\;O \;\;\;\rightarrow  O^{2-} \;+  Cu^{2+} + \mathbf{O}$
\newline
$SrO \;\;:\; Sr^{2+} + 2e^- + \;\;O \rightarrow  Sr^{2+} + [2e^- + O]\; \;\;\;\;\;\;\;\;\;\;\;\;\; \rightarrow  Sr^{2+} + O^{2-}$
\bigskip

 Relative to the host crystal, both the skirmishing and stationary oxygen atoms, $\tilde{O}$ and $O$, appear positive, holding two elementary charges, $+2|e|$, each. 
In this sense, the doping of $La_2CuO_4$ with $Sr$ is often called ``hole doping'' (more accurately, doping the $CuO_2$ planes with holes). 
Both the $\tilde{O}$ and $O$ atoms have a
finite magnetic moment, $\mathbf{m}(\tilde{O}) =\mathbf{m}(O) \ne 0 $, due to their spin quantum number $S = 2\times \frac{1}{2} = 1$ according to the spin configuration $[\uparrow\downarrow]$ $[\uparrow]$ $[\uparrow]$ of their $2p^4$ subshell (Hund's rule of maximal multiplicity). 
The moments of the skirmishing oxygen atoms, $\mathbf{m}(\tilde{O})$---itinerant \textit{via} anion lattice sites---upset the 3D-AFM the host [from $\mathbf{m}(Cu^{2+})$ moments] and cause its collapse at hole density $\tilde{p}=0.02$.

The superlattice, formed by the $O$ ions, gives rise to charge-order stripes. Their incommensurability, in reciprocal lattice units (r.l.u.), depends on $Sr$-doping $x$,
\begin{equation}
q_c(x)  =  \frac{\Omega^{\pm}}{2}\sqrt {x - \tilde{p}} \;,  \;\;\;\; x\le\hat{x}\; ,
\end{equation}
\noindent up to a ``watershed'' doping $\hat{x}$.\cite{6}
The stripe-orientation factor is $\Omega^{+}=\sqrt{2}$ for $x > 0.056$ when stripes are parallel to the $a$ or $b$ axis, but $\Omega^{-} = 1$ for $x < x_6$ when stripes are diagonal. The offset value under the radical is $\tilde{p} \le 0.02$.
A qualitative change of the incommensurability occurs at a watershed concentration of the dopant, $\hat{x}$, which depends on the species of doping and co-doping. It shows up as \emph{kinks} in the $q_c(x)$ profile at $\hat{x}$ (see Fig. 1), where the square-root curve from Eq.(7) levels off to constant plateaus,
\begin{equation}
    q_c(x) = \frac{\sqrt{2}}{2} \sqrt{\hat{x} - \tilde{p}} \;,\;\;\;\; x > \hat{x} \;.
\end{equation}
The charge-order stripes are accompanied by magnetization stripes of incommensurability $q_m(x) = \frac{1}{2} q_c(x)$.
The square-root dependence of $q_c(x)$ results from the spreading of the double holes by Coulomb repulsion to the farthest available separations. 
Thus a rising square-root profile of stripe incommensurability
signifies an underlying superlattice of \emph{lattice-defect charges} (relative to the host crystal).
Increasing density of the doped holes, housing pairwise at lattice-defect $O$ atoms in the $CuO_2$ planes, raises their 
Coulomb repulsion energy. When doping exceeds a watershed value, $x > \hat{x}$, additional holes overflow to the $LaO$ planes\cite{6} where they also reside pairwise in $O$ atoms. This leaves charge-order stripes of \emph{constant} $q_c$ in the $CuO_2$ planes, Eq. (8). 

\includegraphics[width=6.5in]{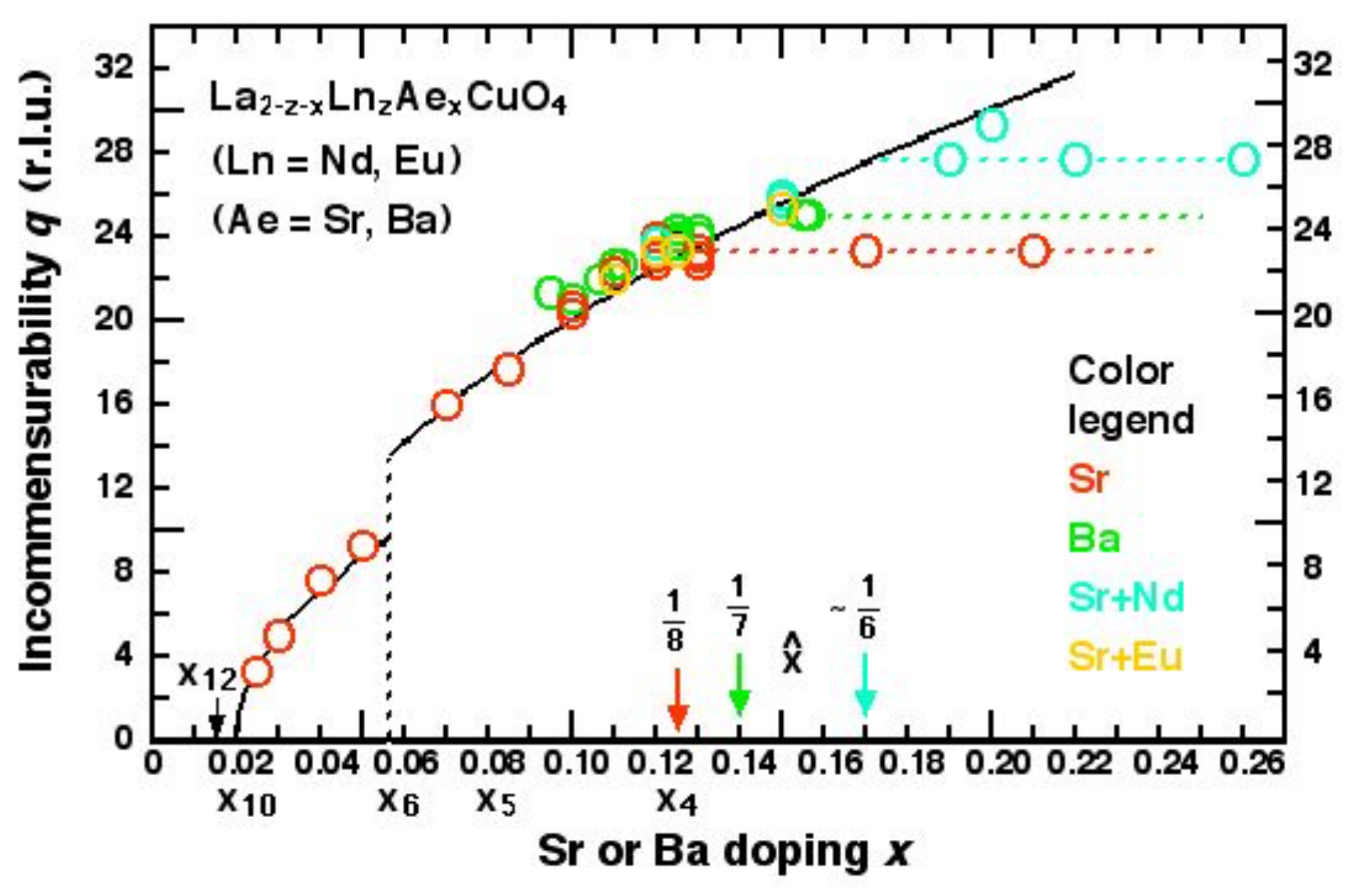}  \footnotesize 

\noindent FIG. 1. Incommensurability of charge-order stripes, $q = q_c$, and of magnetization stripes, $q = 2q_m$, in $La_{2-z-x}Ln_zAe_{x}CuO_{4}$ ($Ln = Nd, Eu; z = 0, 0.4, 0.2$) due to doping with $Ae = Sr$ or $Ba$. Circles show data from X-ray diffraction or neutron scattering. The broken solid curve is a graph of Eq. (7). The discontinuity at $x=0.056$ is caused by a change of stripe orientation from diagonal to parallel, relative to the $Cu$-$O$ bonds. Doping beyond watershed concentrations, $\hat{x}$, yields  constant stripe incommensurabilities, given by Eq. (8) (dashed horizontal lines). \normalsize 

\section{COMMONALITIES IN THE STRIPE AND NQR STUDIES OF $\mathbf{La_{2-x}Sr_xCuO_4}$}

The stripe model assumes that the density of doped holes in the $CuO_2$ plane equals the $Sr$-doping, $p = x$. However, it distinguishes between itinerant holes of density $\tilde{p}$ and stationary holes of density 
$x-\tilde{p}$, located pairwise in the $O$ atoms of the oxygen superlattice. The distinction between both kind of holes is inferred from the collapse of long-range antiferromagnetism (3D-AFM) when $Sr$-doping reaches the N\'{e}el concentration, $x = x_{N0}= 0.02$, at vanishing N\'{e}el temperature, $T_N(x_{N0}) \equiv 0$, and from stripe incommensurability, Eq. (7). As Fig. 1 shows, a large host of data from neutron scattering, hard X-ray diffraction, and resonant soft X-ray scattering is well described by Eq. (7). For low temperatures and low doping ($x < 0.09$), the offset value $\tilde{p}$ in Eq. (7) agrees with the N\'{e}el concentration, $\tilde{p} = x_{N0}= 0.02$.
With more $Sr$ doping, but still at $T\approx0$, it is found that a \emph{smaller} value, $\tilde{p} < x_{N0}$, suffices to keep 3D-AFM suppressed.\cite{6}
Thus the use of $\tilde{p} = 0.02$ in Eq. (7) becomes inaccurate
beyond the low doping range, $x > 0.09$, as it gives \emph{too small} a value for the incommensurabilty $q_{c,m}(x)$. This can be seen in Fig. 1 where in that range most data points cluster slightly above the drawn $q(x)$ curve. 
Use of a \emph{diminished} offset value, $\tilde{p} \simeq 0.015$ in this range (confirmed by recent measurements, as discussed in Ref. 6) shifts that section of the curve slightly upward to better agreement with experiment (not shown). 

Returning to the NQR method, the close agreement of the doped hole density $h$ with $Sr$-doping of $La_{2-x}Sr_xCuO_4$, $h\approx x$, confirms the validity of the approach. Even better is the systematic slight deviation $\Delta h$, Eq. (4), to which no significance may have been attributed previously. As only \textit{stationary} holes would contribute, via EFG, to NQR signals, the NQR method should detect a doped-hole density $x-\tilde{p}$. This strongly suggests the identification $\Delta h = \tilde{p}$, relating to \textit{itinerant} holes. Residing pairwise in skirmishing $\tilde{O}$ atoms, they lead to suppression of 3D-AFM when $x = x_{N0}= 0.02$ and keep 3D-AFM suppressed for $x> x_{N0}$. The values of $x-h$ and $\tilde{p}$ in Table I scatter somewhat about $x_{N0}= 0.02$, possibly due to underlying approximations. (The $x-h$ value for $x=0.24$---from the heavily overdoped range---defies the trend.) 

Since the NQR method\cite{2,3} assumes an \textit{average} hole density $h$ in the $CuO_2$ plane, Eq. 3, it cannot discriminate whether the doped holes reside singularly in $O^-$ ions or doubly in $O$ atoms. The finding from stripe analysis of \textit{double holes} in $O$ atoms awaits confirmation (or refutation) by other experiments.

\section{NQR AND STRIPES IN $\mathbf{n}$-DOPED $\mathbf{Ln_{2-x}Ce_xCuO_4}$,  $\mathbf{Ln = Pr, Nd}$ }  
    
The unit cell of \emph{pristine} ${Ln_2CuO_4}$ ($Ln=Pr,Nd$) has a central $CuO_2$ plane, sandwiched by $LnO$ planes, analogous to ${La_2CuO_4}$.
 Consider again crystal formation by stepwise ionization, where brackets indicate electron localization at atoms, both within the planes and by transfer from the $LnO$ planes to the $CuO_2$ plane:
\bigskip 

\noindent $LnO \;\;:\; Ln^{3+} + 3e^- + \;O \rightarrow  Ln^{3+} + [2e^- + O]\; + \downarrow \overline {e^-\;} | \rightarrow  Ln^{3+} + O^{2-}$
\newline
$CuO_2 :\; Cu^{2+} + 2e^- + 2O \rightarrow  Cu^{2+} + [2e^- + O] \;+ \;\;\;O \;\rightarrow \; O^{2-} \;+  Cu^{2+} + O^{2-}$
\newline
$LnO \;\;:\; Ln^{3+} + 3e^- + \;O \rightarrow  Ln^{3+} + [2e^- + O]\; + 
\uparrow \underline {e^-\;} | \rightarrow  Ln^{3+} + O^{2-}$
\bigskip

\noindent In the simplest case of {\it doping}, $Ce$ substitutes, in some  cells, $Ln$ in both sandwiching planes: 

\bigskip 

\noindent $CeO \;\;:\; Ce^{4+} + 4e^- + \;\;O \rightarrow | \overline {e^-} \downarrow + \; Ce^{4+} + [2e^- + O]\; + 
\downarrow \overline {e^-\;} | \rightarrow  Ce^{4+} + O^{2-}$
\newline
$CuO_2 :\; Cu^{2+} + 2e^- + 2O \rightarrow  Cu^{2+} + 
\;\;\;\;\;\;\;\;\;\;\;\;[2e^- + O] + \;\;\;O \;\;\;\rightarrow    \mathbf{Cu} \;+ \;2O^{2-}$
\newline
$CeO \;\;:\; Ce^{4+} + 4e^- + \;\;O \rightarrow | \underline {e^-} \uparrow +\; Ce^{4+} + [2e^- + O]\; + 
\uparrow \underline {e^-\;} | \rightarrow  Ce^{4+} + O^{2-}$
\bigskip

\noindent Doping lanthanide-based cuprates with cerium, $Ln_{2-x}Ce_xCuO_4$, partially substitutes $Ln^{3+}$  by $Ce^{4+}$, resulting in \emph{electron doping} of the $CuO_2$ plane. As there are \textit{no} itinerant doped electrons in the $CuO_2$ plane, the doped-electron density equals the $Ce$-doping,
\begin{equation}
n = x\;.
\end{equation}
The doped electrons reside pairwise in lattice-site $\mathbf{Cu}$ atoms.\cite{6}
Relative to the host crystal, the $Cu$ atoms have a lattice-defect charge of $-2|e|$. 
Coulomb repulsion spreads the $Cu$ atoms to form a $Cu$ superlattice. Its charge-order incommensurability is given by Eq. (7), with $\tilde{p}=\tilde{n}=0$ (see Fig. 2).

A consequence of smaller ionic radius, $r(Pr^{3+},Nd^{3+})\simeq 1.26$ \AA, compared to $r(La^{3+}) = 1.30$ \AA, the $Ln_{2-x}Ce_xCuO_4$ compounds have the $T'$ crystal structure. It differs from the $T$ 
structure of $La_{2-x}Sr_xCuO_4$ by the position of the $O^{2-}$ ions in the layers that bracket the

\medskip 

\includegraphics[width=5.3in]{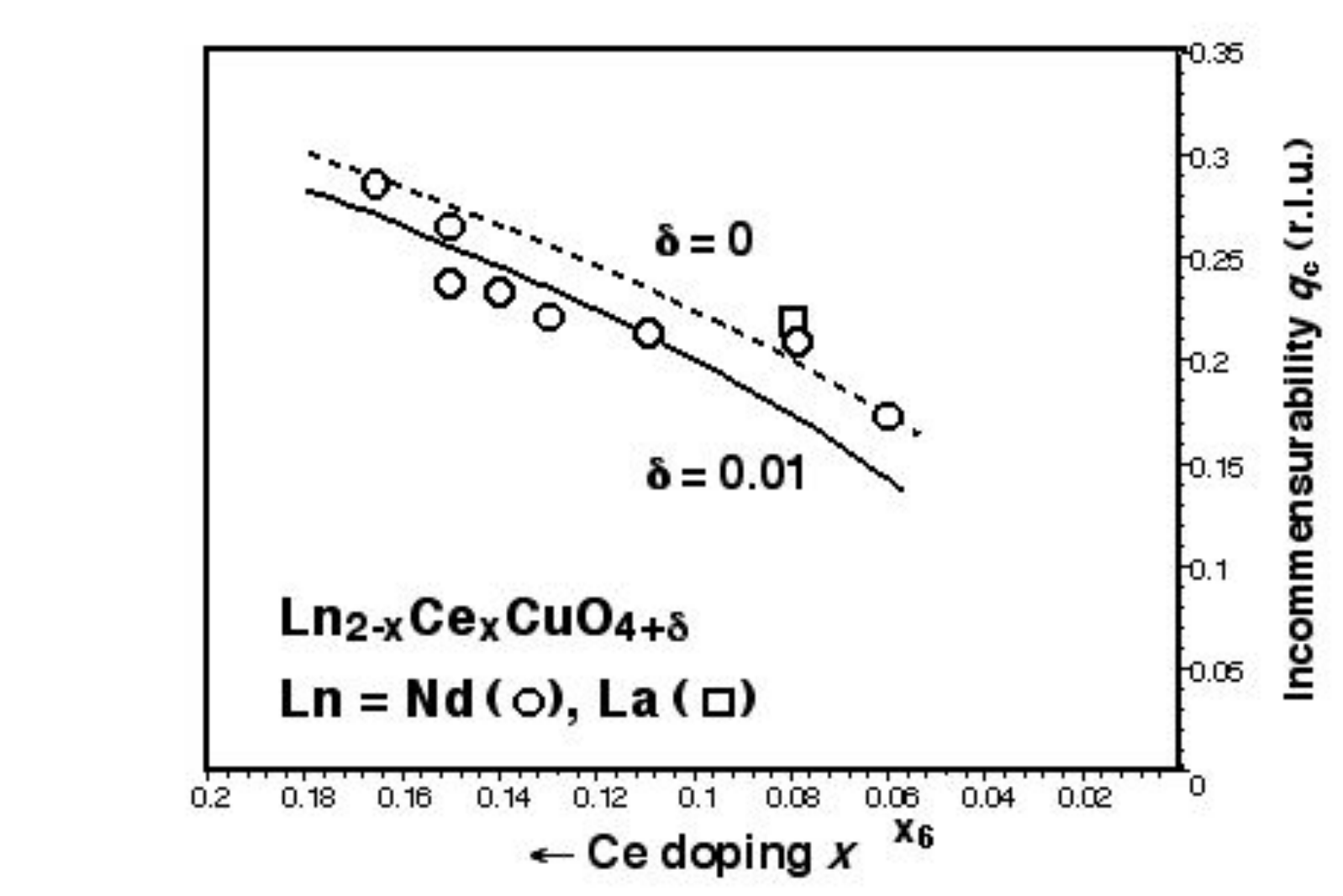}

\footnotesize
\noindent FIG. 2. Observed incommensurability $q_c$ of charge-order stipes in $Ln_{2-x}Ce_xCuO_{4+y}$, $Ln = Nd, La$ (see Ref. 6). The curves are graphs of Eq. (7) without excess oxygen ($\delta=0$, dashed line) and with excess oxygen of $\delta=0.01$ (solid line). 
\normalsize

\noindent $CuO_2$ plane. In the $T$ structure those $O^{2-}$ ions are at apical positions (above or beneath 
the $Cu^{2+}$ ions) whereas in the $T'$ structure they reside above or beneath $O^{2-}$ ions of the $CuO_2$ plane. As a result, the unit cell of the parent crystal $Ln_2CuO_4$ is wider and shorter  ($a\simeq3.95$ \AA, $c\simeq12.19$ \AA) than of $La_2CuO_4$ ($a=3.81$ \AA, $c=13.2$ \AA).\cite{8}

For reasons of stability, $Ln_{2-x}Ce_{x}CuO_4$ crystals need to be grown with excess oxygen, $O_\delta$, to be eliminated in post-growth annealing.\cite{8} 
 As Eq. (7) is based on Coulomb repulsion of like charges in the $CuO_2$ planes, its success for stripes in electron-doped `214' compounds (of $T'$ structure) implies that the excess oxygen must reside
 interstitially as oxygen \textit{ions}, $O_\delta^{2-}$, in or between the $LnO$ planes.\cite{6} Any residual excess oxygen results in a \emph{hole}-doping contribution to the $CuO_2$ planes that correspondingly reduces their electron-doping from $Ce$. Accordingly, the charge-order incommensurability of $Ln_{2-x}Ce_{x}CuO_{4+\delta}$ can be expressed with Eq. (7) using $\tilde{p} = 2\delta$.\cite{6}

The NQR investigations of $Ln_{2-x}Ce_xCuO_4$ yield doped-electron densities very close to the $Ce$-doping, approximating Eq. (9).\cite{4,5} It is found that doping with electrons predominantly decreases $n_d$ but only slightly $n_p$. In other words, the doped electrons in $Ln_{2-x}Ce_xCuO_4$ reside almost entirely at $Cu$ atoms.
The copper signal $\nu_{Cu}$ is very sensitive to $Ce$-doping, having a wide background of broadened satellites such that the signal effectively disappears at $x=0.15$.
For oxygen, the NQR analysis obtains a hole content $n_p \simeq 0.20$ in the parent $Ln_2CuO_4$ compounds, compared to $n_p \simeq 0.10$ in $La_2CuO_4$. 

A comparative interpretation of the NQR results from hole-doped and electron-doped cuprates is complicated by several factors:\cite{9,10,11} (i) the EFG at a copper or oxygen nucleus from a hole in its own electron shell (here called ``self EFG''); (ii) the EFG from differently charged nearest neighbors in the $CuO_2$ plane (abreviated ``nn EFG''); (iii) the presence or absence of $O^{2-}$ ions above and beneath $Cu^{2+}$ ions---called ``apical'' oxygen ions---and of $O^{2-}$ above and beneath $O^{2-}$ ions in the $CuO_2$ plane according to the $T$ or $T'$ structure; (iv) hybridization of $Cu\;3d_{x^2-y^2}$ and $O\;2p_\sigma$ orbitals leading to covalent bond, (v) deformation of electron shells in the crystal, and (vi) higher-order effects.

Using the finding from the stripe model that doped holes and electrons are hosted \textit{pairwise} in $O$ and $Cu$ atoms, respectively, the following diagrams provide a qualitative overview of possible cases with combinations of the influences (i), (ii) and (iii). The top diagrams show planar configurations of the copper sites and the bottom diagrams those of the oxygen sites. The middle panels give the configurations in the parent crystals and the outer panels those of the doped sites. The largest effect on NQR can be expected from self EFG, noted beneath the corresponding diagrams. 
(No self EFG exists at the $Cu$ nucleus due to the spherical $4d^{10}5s^1$ orbitals of the atom.)
Noticeable effects can also be expected from planar nn EFG and from the presence or absence of nearest neighbors in the sandwiching layers according to the compounds' $T$ or $T'$ structure. 
Here the \textbf{absence} of $O^{2-}$ above and beneath $\mathbf{Cu^{2+}}$ and $\mathbf{Cu}$ in $Pr_2CuO_4$ and $Pr_{2-x}Ce_xCuO_4$, respectively, is marked bold in the diagrams $(b,a)$.
Likewise, the \textbf{absence} of $O^{2-}$ ions above and beneath $\mathbf{O^{2-}}$ and, respectively, $\mathbf{O}$ in the $CuO_2$ plane of $La_2CuO_4$ is marked bold in $(g,h)$.  
In all cases the effects of undoped and doped sites are different. Thus it is not surprising that the corresponding NQR frequencies are different. This renders a comparison difficult.

\bigskip 

\noindent .$\;\;$electron doped$\;\;\;\;\;\;\;\;\;\;\;\;$parent   $(b,f)\;\;\;\;\;\;\;\;\;$parent$(c,g)\;\;\;\;\;\;\;\;\;\;$ hole doped$\;\;\;\;$.
.$\;\;Pr_{2-x}Ce_xCuO_4\;\;\;\;\;\;\;\;\;\;\;\;\;\;\;\;\;\;\;Pr_2CuO_4\;\;\;\;\;\;\;\;\;\;\;\;\;\;\;\;\;\;\;La_2CuO_4\;\;\;\;\;\;\;\;\;\;\;\;\;\;\;\;\;\;\;La_{2-x}Sr_xCuO_4\;\;$.

\bigskip

\noindent .$(a)\;\;\;\;\;\;\;\;\;\;O^{2-}\;\;\;\;\;\;\;\;\;\;\;\;\;\;\;\;(b)\;\;\;\;\;\;\;\;\;O^{2-}\;\;\;\;\;\;\;\;\;\;\;\;\;\;\;\;\;(c)\;\;\;\;\;\;\;\;\;O^{2-}\;\;\;\;\;\;\;\;\;\;\;\;\;\;\;\;\;\;\;\;\;(d)\;\;\;\;\;\;\;\;\;\;O^{2-}\;\;\;\;\;\;\;\;\;\;\;\;\;\;\;\;$.
\newline  .$\;\;\;\;\;O^{2-}\;\;\;\mathbf{Cu}\;\;\;O^{2-}\;\;\;\;\;\;\;\;\;\;\;\;\;O^{2-}\;\;\mathbf{Cu^{2+}}\;\;O^{2-}\;\;\;\;\;\;\;\;\;\;\;O^{2-}\;\;Cu^{2+}\;\;O^{2-}\;\;\;\;\;\;\;\;\;\;\;\;\;\;\;\;O^{2-}\;\;\;Cu^{2+}\;\;\;O$
\newline  .$\;\;\;\;\;\;\;\;\;\;\;\;\;\;\;O^{2-}\;\;\;\;\;\;\;\;\;\;\;\;\;\;\;\;\;\;\;\;\;\;\;\;\;\;\;\;\;O^{2-}\;\;\;\;\;\;\;\;\;\;\;\;\;\;\;\;\;\;\;\;\;\;\;\;\;\;\;\;\;\;\;O^{2-}\;\;\;\;\;\;\;\;\;\;\;\;\;\;\;\;\;\;\;\;\;\;\;\;\;\;\;\;\;\;\;\;\;\;\;\;O^{2-}$
\newline .$\;\;\;\;\;\;\;\;\;\;\;\;\;\;\;\;\;\;\;\;\;\;\;\;\;\;\;\;\;\;\;\;\;\;\;\;\;\;\;\;\;\;\;\;\;\;$self EFG$\;\;\;\;\;\;\;\;\;\;\;\;\;\;\;\;\;\;\;\;\;\;\;$self EFG$\;\;\;\;\;\;\;\;\;\;\;\;\;\;\;\;\;\;\;\;\;\;\;\;\;\;\;$self EFG
\newline .$\;\;\;\;\;\;\;\;\;\;\;\;\;\;\;\;\;\;\;\;\;\;\;\;\;\;\;\;\;\;\;\;\;\;\;\;\;\;\;\;\;\;\;\;\;\;\;\;\;\;\;\;\;\;\;\;\;\;\;\;\;\;\;\;\;\;\;\;\;\;\;\;\;\;\;\;\;\;\;\;\;\;\;\;\;\;\;\;\;\;\;\;\;\;\;\;\;\;\;\;\;\;\;\;\;\;\;\;\;\;\;\;\;\;\;\;\;\;\;\;$ + nn EFG

\bigskip 
\noindent .$(e)\;\;\;\;\;\;\;\;Cu^{2+}\;\;\;\;\;\;\;\;\;\;\;\;\;\;\;\;(f)\;\;\;\;\;\;\;Cu^{2+}\;\;\;\;\;\;\;\;\;\;\;\;\;\;\;\;\;(g)\;\;\;\;\;\;\;Cu^{2+}\;\;\;\;\;\;\;\;\;\;\;\;\;\;\;\;\;\;\;\;\;(h)\;\;\;\;\;\;\;\;Cu^{2+}\;\;\;\;\;\;\;\;\;\;\;\;\;\;\;\;$.
\newline  .$\;\;\;\;\;\;Cu\;\;\;O^{2-}\;Cu^{2+}\;\;\;\;\;\;\;\;\;\;Cu^{2+}\;\;O^{2-}\;\;Cu^{2+}\;\;\;\;\;\;\;\;\;\;Cu^{2+}\;\;\mathbf{O^{2-}}\;\;Cu^{2+}\;\;\;\;\;\;\;\;\;\;\;\;\;\;\;\;Cu^{2+}\;\;\mathbf{O}\;\;\;Cu^{2+}$
\newline  .$\;\;\;\;\;\;\;\;\;\;\;\;\;Cu^{2+}\;\;\;\;\;\;\;\;\;\;\;\;\;\;\;\;\;\;\;\;\;\;\;\;\;\;\;\;Cu^{2+}\;\;\;\;\;\;\;\;\;\;\;\;\;\;\;\;\;\;\;\;\;\;\;\;\;\;\;\;\;Cu^{2+}\;\;\;\;\;\;\;\;\;\;\;\;\;\;\;\;\;\;\;\;\;\;\;\;\;\;\;\;\;\;\;\;\;Cu^{2+}$
\newline .$\;\;\;\;\;\;\;\;\;\;\;$nn EFG$\;\;\;\;\;\;\;\;\;\;\;\;\;\;\;\;\;\;\;\;\;\;\;\;\;\;\;\;\;\;\;\;\;\;\;\;\;\;\;\;\;\;\;\;\;\;\;\;\;\;\;\;\;\;\;\;\;\;\;\;\;\;\;\;\;\;\;\;\;\;\;\;\;\;\;\;\;\;\;\;\;\;\;\;\;\;\;\;\;\;\;\;\;\;\;\;\;\;\;\;$self EFG

\section{NQR DETERMINATION OF DOPED-HOLE DENSITY IN $\mathbf{YBa_2Cu_3O_{6+y}}$}

Using the same procedure as for $La_{2-x}Sr_xCuO_4$, Haase \textit{et al.}\cite{2} obtain for $YBa_2Cu_3O_{6+y}$ the doped-hole densities $h$ listed in Table II. The results are comparable with values from the universal-dome method and from stripe incommensurability.\cite{1,7} This confirms again the validity of the NQR approach to determine $h$.
However, in striking contrast to the heterovalent-metal-doped compounds, $La_{2-x}Sr_xCuO_4$ and $Ln_{2-x}Ce_xCuO_4$
($Ln = Pr, Nd$), 
where the doped holes and electrons were found to reside almost entirely at oxygen and, respectively,
copper ions, distinctly different orbital occupancies are found with NQR for oxygen-doped 

\begin{table}[ht!]
\caption{Doped-hole density $h$ in $YBa_2Cu_3O_{6+y}$, obtained from NQR data for copper $d$ and oxygen $p_c$ orbitals,\cite{2} from stripe incommensurability,\cite{7} and from the universal dome method.\cite{1}}
\begin{tabular}{|p{1.15cm}||p{2.1cm}|p{2.3cm}|p{3.7cm}|p{2cm}|p{2cm}|p{2cm}|p{2cm}|}
 \hline  \hline
$\;\;\;y$&$\;h \leftarrow$ NQR &$\;h \leftarrow$ Stripes& $\;h$ $\leftarrow$ Universal Dome\\
 \hline  \hline
$\;0.00$&&&$\;\;\;\;\;\;\;\;\;\;\;\;\;0.00$ \\  
$\;0.31$&&$\;\;\;\;\;\;\;0.00$&$\;\;\;\;\;\;\;\;\;\;\;\;\;0.05$ \\  
$\;0.45$&&$\;\;\;\;\;\;\;0.07$&$\;\;\;\;\;\;\;\;\;\;\;\;\;0.08$  \\
$\;0.50$&$\;\;\;\;\;\;0.00$&$\;\;\;\;\;\;\;0.09$&$\;\;\;\;\;\;\;\;\;\;\;\;\;0.08$ \\
$\;0.60$&$\;\;\;\;\;\;0.10$&$\;\;\;\;\;\;\;0.12$&$\;\;\;\;\;\;\;\;\;\;\;\;\;0.09$ \\
$\;0.63$&$\;\;\;\;\;\;0.11$&$\;\;\;\;\;\;\;0.13$&$\;\;\;\;\;\;\;\;\;\;\;\;\;0.10$\\
$\;0.68$&$\;\;\;\;\;\;0.14$&$\;\;\;\;\;\;\;0.14$&$\;\;\;\;\;\;\;\;\;\;\;\;\;0.12$ \\
$\;0.75$&&$\;\;\;\;\;\;\;0.16$&$\;\;\;\;\;\;\;\;\;\;\;\;\;0.14$  \\
$\;0.92$&&$\;\;\;\;\;\;\;0.18$&$\;\;\;\;\;\;\;\;\;\;\;\;\;0.16$  \\
$\;0.96$&$\;\;\;\;\;\;0.16$&&$\;\;\;\;\;\;\;\;\;\;\;\;\;0.19$  \\
 \hline   \hline
\end{tabular}

\label{table:2}  \end{table}

\noindent and oxygen-enriched compounds.\cite{3,4,5} 
With Eqs. (5, 6) one can obtain the orbital probabilities $P_d$ and $P_p$, according to which a doped hole resides at a copper or oxygen ion in the $CuO_2$ plane, by the slope,
\begin{equation}
    s = \frac{n_d - n_{d0}}{2n_p - 2n_{p0}} = \frac{P_d}{P_p} \;,
\end{equation}
of orbital-occupation data in an $n_d$ \textit{vs.} $2n_p$ display. Together with $P_d + P_p  \equiv 1$, this gives 
\begin{equation}
    P_d = \frac{s}{s+1}\;\;\;\;\;and\;\;\;\;\;    P_p = \frac{1}{s+1} \;.
\end{equation}
In Fig. 5 of Ref. 4  the $n_d$ \textit{vs.} $2n_p$ values fall along straight lines for compounds families. Their slopes are listed
in Table III along with a doped hole's orbital-occupation probabilities. In oxygen-doped $YBa_2Cu_3O_{6+y}$ and $YBa_2Cu_4O_8$,  $P_d \approx \frac{1}{3}$ and $P_p \approx \frac{2}{3}$ is found. In oxygen-enriched $Hg$-, $Bi$-, and $Tl$-based cuprates, they are essentially equal, $P_d \simeq P_p \simeq \frac{1}{2}$. How can these findings be understood and what do they tell us?

\section{STRIPES IN $\mathbf{YBa_2Cu_3O_{6+y}}$}

The unit cell of $YBa_2Cu_3O_{6+y}$ has two $CuO_2$ planes, separated by the $Y$ plane and bracketed by $BaO$ layers. At the top (and bottom) of each unit cell of \textit{undoped} $YBa_2Cu_3O_6$ 

\pagebreak 

\begin{table}[ht!]
\caption{Slopes $s$ of orbital-occupation data from Ref. 4, Fig. 5, and occupation probabilities $P_d$ and $P_p$ of doped copper $d$ and oxygen $p_{\sigma}$ orbitals for heterovalent-metal-doped cuprates (upper part) and oxygen-doped/enriched cuprates (lower part).}
\begin{tabular}{|p{10.5cm}|p{1.5cm}||p{1.5cm}|p{1.5cm}|}
 \hline  \hline
Compound&$\;\;\;\;s$& $\;\;\;P_d$& $\;\;\;P_p$\\
 \hline  \hline
$La_{2-x}Sr_xCuO_4$&$\;\;\;0.14$&$\;\;\;0.12$&$\;\;\;0.88$\\ \hline
$Pr_{2-x}Ce_xCuO_4$&$\;\;\;4.82$&$\;\;\;0.83$&$\;\;\;0.17$\\
$Nd_{2-x}Ce_xCuO_4$&$\;\;\;6.50$&$\;\;\;0.87$&$\;\;\;0.13$\\ \hline \hline
$YBa_2Cu_3O_{6+y}\;,\;YBa_2Cu_4O_8$&$\;\;\;0.55$&$\;\;\;0.36$&$\;\;\;0.64$\\ \hline
$HgBa_2CuO_{4+\delta}\;,\;Bi_2Ba_2CaCu_2O_{8+\delta} \;,$&$\;\;\;0.96$&$\;\;\;0.49$&$\;\;\;0.51$\\
$Tl_2Ba_2CuO_{6+\delta}\;,\;Tl_2Ba_2CaCu_2O_{8+\delta}  \;,\; 
Tl_2Ba_2Ca_2Cu_3O_{10+\delta}$&$\;\;\;0.96$&$\;\;\;0.49$&$\;\;\;0.51$\\
\hline   \hline
\end{tabular}
\label{table:3}  \end{table}

\noindent is a plane of $Cu^+$ ions (called the ``$Cu(1)$ plane''). In order to introduce the effects of doping we temporarily make a simplifying assumption (here called the ``0.5-watershed''), that we'll later drop when more familiar. By this approximation, oxygen doping $y\le 0.5$ causes ionization in the $Cu(1)$ plane, $O + 2Cu^+ \rightarrow 2Cu^{2+} + O^{2-}$, with the $O^{2-}$ ions residing between $Cu^{2+}$ ions along the crystal's $b$ direction, called $CuO$ chains. The $CuO_2$ planes remain unaffected by this doping. Accordingly \textit{no} charge-order stripes are observed for $y\le0.5$. The filling of $Cu^{2+}O^{2-}$ chains is completed at doping $y=0.5$, as can be seen by stoichiometry and charge balance of $Y^{3+}Ba^{2+}_2Cu^{2+}_3O^{2-}_{6.5}$. The $Cu(1)$ plane of $YBa_2Cu_3O_{6.5}$ alternates with filled and empty $CuO$ chains (called ``ortho-II oxygen ordered'').

By the 0.5-watershed approximation, doping $y>0.5$ causes incorporation of $O$ \textit{atoms} either in more $CuO$ chains of the $Cu(1)$ plane or in the $CuO_2 \equiv Cu(2)$  planes,
\begin{equation}
y = \chi^{Cu(1)}(y) + 2\delta^{Cu(2)}(y)\;.
\end{equation}
It is the density of embedded oxygen, $\delta^{Cu(2)}\equiv\delta$, in each $CuO_2$ plane that gives rise to charge-order stripes. Their incommensurability [averaged over stripes along the $a$ and $b$ direction, $\overline{q}_c = (q_c^a + q_c^b)/2$],
is given by the empirical formula\cite{7}
\begin{equation}
\overline{q}_c(y) = \overline{q}_c(y_c^{ons}) - \gamma \times [\delta(y) - \delta(y_c^{ons})]\;.
\end{equation}
Here the superscript ‘\textit{ons}’ indicates the onset of charge-order (\textit{c}) stripes. The onset incommensurability is material-specific:  $\overline{q}_c(y_c^{ons}) = 0.33$ for $YBa_2Cu_3O_{6+y}$. Both the coefficient $\gamma = 0.64$/r.l.u. and the onset density $\delta(y_c^{ons})=0.035$ hold for the whole family of oxygen-

\includegraphics[width=5.9in]{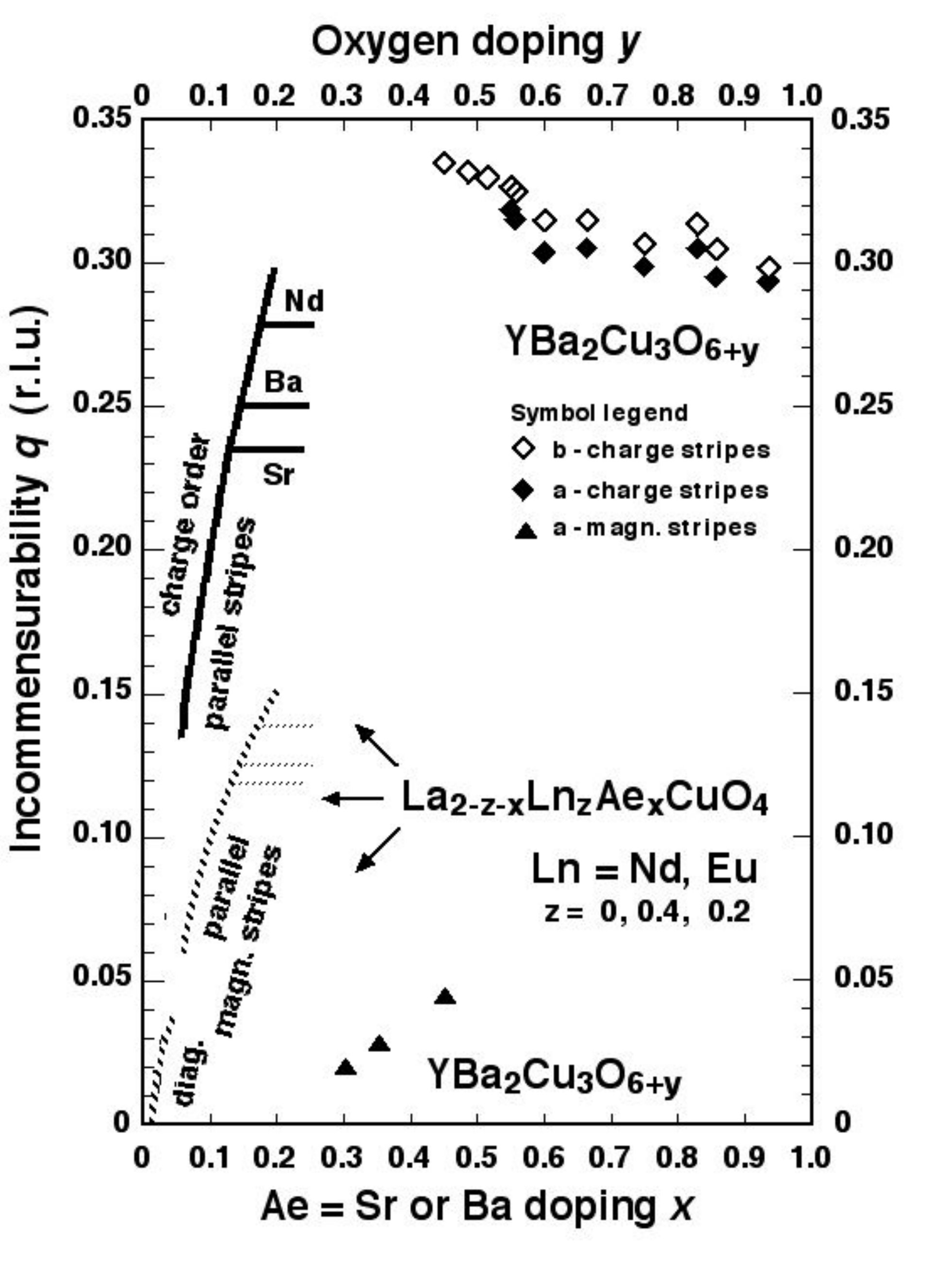}

\footnotesize 

\noindent FIG. 3. Incommensurability of charge-order stripes (solid line) and magnetization stripes (hatched line) in $La_{2-z-x}Ln_zAe_{x}CuO_{4}$ of the `214' family due to $Ae$ doping with $Ae = Sr$ or $Ba$ (curves on the left, equivalent to Fig. 1) and in oxygen-doped $YBa_2Cu_3O_{6+y}$ near the top (and near the bottom, if insufficiently annealed).
\normalsize 

\noindent doped/enriched cuprates.\cite{7} Figure 3 shows the observed incommensurabilties of charge-order stripes $(q_c^a$ and $q_c^b)$, in $YBa_2Cu_3O_{6+y}$ (and $q_m^a$ of magnetization stripes if the crystals are not sufficiently annealed---not discussed here). 
For comparison, the (re-scaled) result from Fig. 1 is
included in Fig. 3 on the left. 
The doping dependence of charge-order stripes is distinctly different in the alkaline-earth-doped cuprates, $La_{2-x}Ae_xCuO_4$ ($Ae = Sr, Ba$), and in the oxygen-doped cuprate, $YBa_2Cu_3O_{6+y}$, showing a square-root dependence of the former, Eq. (7), but almost constant values of the latter, Eq. (13). The square-root dependence results from Coulomb repulsion of the doped holes in the $CuO_2$ plane (residing pairwise in lattice-site $O$ atoms) to the largest separation, causing formation of an $O$ superlattice. Conversely, the almost constant $q_c(y)$ values of $YBa_2Cu_3O_{6+y}$ result from the implementation of \textit{neutral} $O$ atoms in the $CuO_2$ planes. (The slight decrease of $q_c(y)$ with increasing $O$-doping $y$ is caused by secondary effects.\cite{7}) 
The distinctly different doping dependence of charge-order incommensurability disproves the common misconception that doped oxygen ionizes in $YBa_2Cu_3O_{6+y}$ as $O_y \rightarrow O^{2-} + 2e^+$ and separates---with $O^{2-}$ entering the so-called charge-reservoir layer (that is, all planes except $CuO_2$) and the holes, $2e^+$, entering the $CuO_2$ planes. If this were true, then a rising square-root doping dependence of charge-order stripes in $YBa_2Cu_3O_{6+y}$ would result, \textit{contrary} to the observed $q_c(y) \approx const$.\cite{7}

It has been proposed\cite{7} that each doped oxygen atom is embedded in the $CuO_2$ plane at an interstitial site between four $O^{2-}$ ions, called ``pore'' (see Fig. 4). The oxygen atom at the pore site, denoted $\mathring{O}$, bonds with two $O^{2-}$ neighbors (abbreviated as $\ddot{O} \equiv O^{2-}$) to form an ozone molecule ion, $\ddot{O}\mathring{O}\ddot{O}$.
Linked by an intermediate oxygen ion each (marked bold, $\mathbf{\ddot{O}}$), the ozone molecules line up along the crystal's $a$ or $b$ direction to form trains of $\ddot{O}\mathring{O}\ddot{O}\mathbf{\ddot{O}}$ motifs that are observed as charge-order stripes in the $CuO_2$ planes of oxygen-doped $YBa_2Cu_3O_{6+y}$ and $YBa_2Cu_4O_8$, as well as of oxygen-enriched, $HgBaCuO_{4+\delta}$, and the $Bi$- and $Tl$-based cuprates.\cite{7} The distinction of oxygen-\textit{doping} $y$, and oxygen-\textit{enrichment} $\delta$, signifies the stoichiometric exactitude of $y$, in contrast to the uncertainty of $\delta$ due to diffusive procedures and thermal treatment. 
For $YBa_2Cu_3O_{6+y}$, the doping certainty is of little help, however, when it comes to the oxygen content of the $CuO_2$ planes as the doped oxygen is shared to an uncertain degree with the $CuO$ chains, Eq. (12).

The ``0.5-watershed'' simplification, employed for ease of introduction, holds \textit{approximately}, but not strictly: To a small degree, $O$ atoms are already embedded in the $CuO_2$ planes for $y<0.5$---weak charge-order stripes are already observed in the doping range 

\includegraphics[width=6.2in]{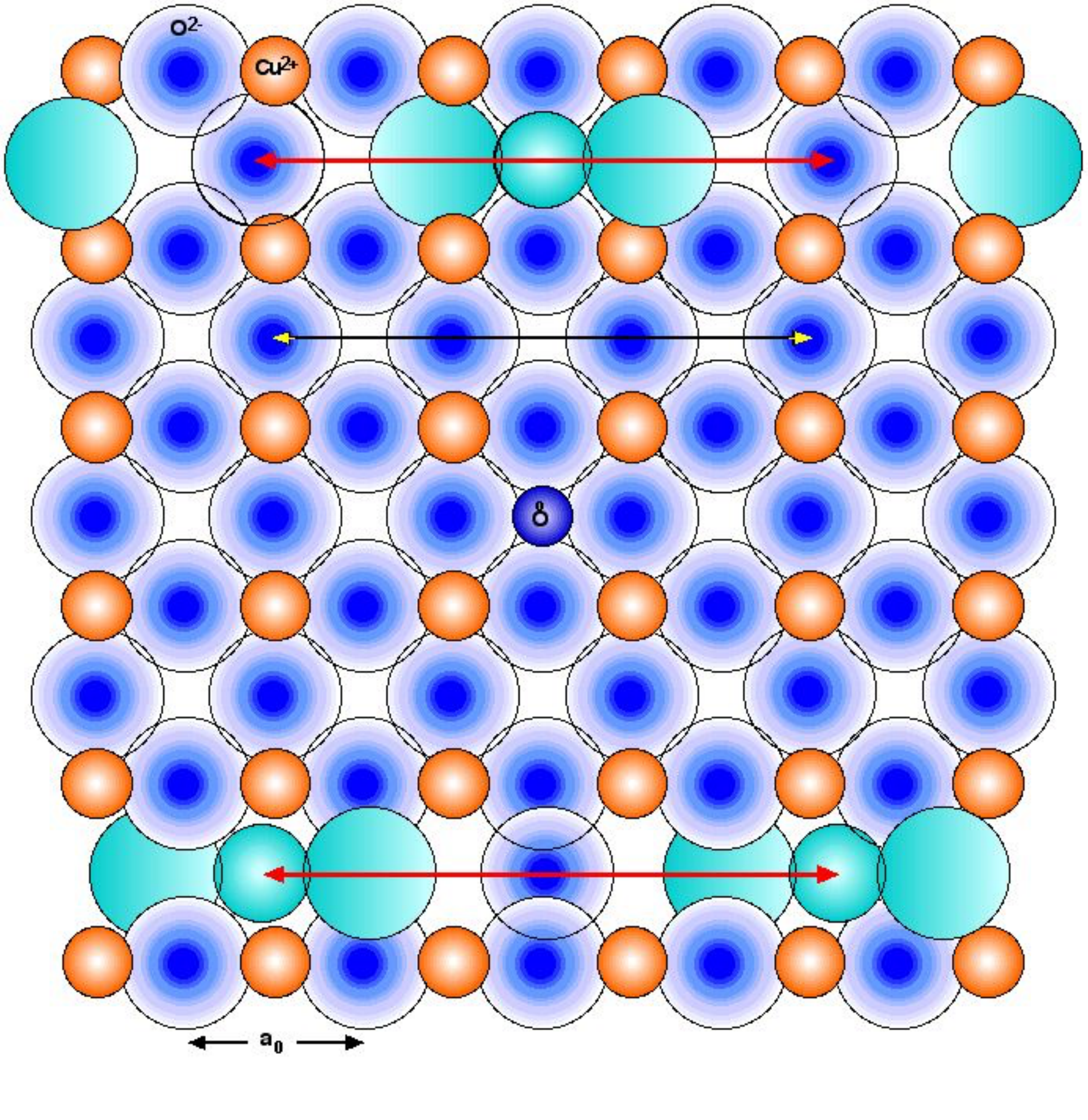}

\footnotesize 

\noindent FIG. 4. Doped oxygen in the $CuO_2$ plane of $YBa_2Cu_3O_{6+y}$ ($y \simeq 0.55$), shown schematically for an oxygen atom at the stages of embedding (center, sixth row) and final relaxation (second row top and bottom). In the space (``pore'') between four $O^{2-}$ ions (denoted $\ddot{O}$ for short), the embedded oxygen atom (denoted $\mathring{O}$) bonds with two $O^{2-}$ neighbors, here in the $a$-direction, to form an ozone molecule ion, $\ddot{O}\mathring{O}\ddot{O}$ (turquoise color).
Linked by an intermediate oxygen ion each (marked bold, $\mathbf{\ddot{O}}$), the ozone molecules line up to form trains of $\ddot{O}\mathring{O}\ddot{O}\mathbf{\ddot{O}}$ motifs. The length of a motif is $L \simeq 3.125\; a_0$ (red double arrow), slightly larger than $3a_0$ (black double arrow). Its reciprocal gives the incommensurabiliy of charge-order stripes, $q_c =1/L \simeq 0.32$ r.l.u. 
\normalsize 

\pagebreak
\noindent $0.45<y<0.5$, and low-$T_c$ superconductivity in $0.31<y<0.5$. Accordingly, some $CuO$ chain filling with $O^{2-}$ ions continues for $y>0.5$ besides the then prevalent filling with $O$ atoms.

\section{COMPARISON OF STRIPE AND NQR STUDIES OF HIGH-$\mathbf{T_c}$ CUPRATES}

The relevant properties of heterovalent-metal-doped and oxygen-enriched cuprates are given in Table IV. The hole-doped example of $La_{2-x}Sr_xCuO_4$ suffices for the comparison, as electron-doped cases correspond equivalently (see table caption).
In order to avoid complications from $CuO$ chain filling 
in oxygen-doped $YBa_2Cu_3O_{6+y}$, we compare with oxygen-enriched $HgBa_2CuO_{4+\delta}$, where \textit{all} excess oxygen is embedded in the $CuO_2$ plane.

The outstanding distinction is a hole-doping of the $CuO_2$ plane of $La_{2-x}Sr_xCuO_4$, with holes residing pairwise at lattice-site $O$ atoms, but of neutral $O$-atom-embedding in $HgBa_2CuO_{4+\delta}$, located at interstial positions (``pores''). Strictly speaking, calling the 

\begin{table}[ht!]
\begin{tabular}{|p{7cm}|p{5cm}|p{4cm}|}
 \hline  \hline
Property&$\;\;\;\;\;\;\;\;\;\;\;\;$hole-doped&$\;\;\;\;\;$oxygen-enriched\\
 \hline  \hline
Example compound&$\;\;\;\;\;\;\;\;\;\;La_{2-x}Sr_xCuO_4$&
$\;\;\;\;\;\;\;HgBa_2CuO_{4+\delta}$\\ \hline
Doping/enrichment of the crystal with& $\;\;\;\;\;\;\;\;\;\;\;\;\;\;\;\;\;\;\;\;Sr$&$\;\;\;\;\;\;\;\;\;\;\;\;\;\;\;\;\;O$\\
causes doping of the $CuO_2$ plane with& $\;\;\;\;\;\;\;\;\;\;\;\;\;\;\;\;\;\;\;\;\;e^+$&$\;\;\;\;\;\;\;\;\;\;\;\;\;\;\;\;\;\mathring{O}$\\
Net charge accumulation in $CuO_2$ plane? &$\;\;\;\;\;\;\;\;\;\;\;\;\;\;\;\;\;\;\;\;$yes&$\;\;\;\;\;\;\;\;\;\;\;\;\;\;\;\;\;$no\\ \hline
From stripe analysis:&$\;\;\;\;\;\;\;\;\;\;2e^++ \underline{O}^{2-} \rightarrow \underline{O}$&$\;\;\;\;\;\;\;\;\;\;\;\;\;O \rightarrow \mathring{O}$\\ 
Location of doping-affected oxygen&$\;\;\;\;\;\;\;\;\;\;\;\;$at \underline{lattice site}&$\;$at interstitial position\\  \hline
Defect charge of doping-affected oxygen&$\;\;\;\;\;\;\;\;\;\;\;\;\;\;\;\underline{Q} = 2|e|$&$\;\;\;\;\;\;\;\;\;\;\;\;\;\mathring{Q}=0$\\
Coulomb repulsion of Q's?&$\;\;\;\;\;\;\;\;\;\;\;\;\;\;\;\;\;\;\;\;$yes&$\;\;\;\;\;\;\;\;\;\;\;\;\;\;\;\;\;$no\\ \hline
Stripe incommensurability&$\;\;\;\;\;\;\;\;\;\;q_c(x) \propto \sqrt{x-\tilde{p}}$&$\;\;\;\;\;\;\;\;q_c(\delta)\approx const.$\\ \hline
NQR: doped orbital probability&$\;\;\;\;\;\;\;\;\;P_d\approx \frac{1}{8}, \;P_p\approx \frac{7}{8}$&$\;\;\;\;\;\;\;\;P_d \simeq \;P_p \simeq \frac{1}{2}$\\ \hline
NQR interpretation:& doped holes $\rightarrow$ latt.-site oxy.&$\;\;\;\;\;\;\;\;\;\;\;\;\;\;\;\;\;\;$ ?\\ 
 \hline   \hline
 \end{tabular}
 \caption{Comparison of hole-doped and oxygen-enriched cuprates.
Electron-doped cases correspond equivalently to hole-doping with replacements 
$La_{2-x}Sr_xCuO_4 \rightarrow Pr_{2-x}Ce_xCuO_4; 
Sr \rightarrow Ce$; 
$e^+ \rightarrow e^-$; $2e^++ \underline{O}^{2-} \rightarrow \underline{O}\;\Rightarrow \; 2e^-+ \underline{Cu}^{2+} \rightarrow \underline{Cu}$; $\underline{Q} \rightarrow -2|e|;
P_d \rightarrow \frac{6}{7}, \;P_p\rightarrow \frac{1}{6}$; doped electrons $\rightarrow$ lattice-site copper.}
\label{table:4}  \end{table}

\noindent presence of excess oxygen, $O_{\delta}$, in the $CuO_2$ plane ``hole doping'' is a misnomer---``oxygen-atom doping'' would be a more accurate (although awkward) term. Despite their common lack of two electrons from a closed $2p^6$ shell, lattice-site $\underline{O}$ atoms and interstitial $\mathring{O}$ atoms have different properties due to their different position in the crystal. In $La_{2-x}Sr_xCuO_4$, Coulomb  repulsion spreads the effective defect-charges of the lattice-site oxygen, $\underline{Q} = 2|e|$, to an $O$ superlattice which gives rise to charge-order stripes. Their incommensurability has a rising square-root dependence on $Sr$ doping $x$. 
In contrast, the excess $\mathring{O}$ atoms in $HgBa_2CuO_{4+\delta}$, lacking a net defect charge, $\mathring{Q} = 0$, are \textit{not} spread by Coulomb force but aggregate to $\ddot{O}\mathring{O}\ddot{O}\mathbf{\ddot{O}}$ trains. They give rise to charge-order stripes with with almost constant incommensurability. 
The doped orbital probabilities $P_{d,p}$ from NQR confirm that
the doped holes in $La_{2-x}Sr_xCuO_4$ reside almost entirely at lattice-site oxygen. Likewise, the doped electrons in $Pr_{2-x}Ce_xCuO_4$ are found by NQR to reside almost entirely at lattice-site copper. 

This brings us to the doped-orbital probabilities in $HgBa_2CuO_{4+\delta}$, $P_d\simeq P_p\simeq 0.5$.
The NQR formalism by Haase \textit{et al.} assigns doped holes to \textit{lattice-site} ions, Eqs. (1, 2), but makes no allowance for doped $O$ atoms at interstitial sites of the $CuO_2$ plane.
The following diagram shows a doped oxygen atom at the pore position, $\mathring{O}$, and its planar environment. 

\bigskip 
\noindent .$\;\;\;\;\;\;\;\;\;\;\;\;\;\;\;\;\;\;\;\;\;\;\;\;\;\;\;\;\;
\;\;\;\;\;O^{2-}\;\;\;\;\;\;\;\;\;\;\;\;O^{2-}\;\;\;\;\;\;\;\;\;\;\;\;\;O^{2-}\;\;\;\;\;\;\;\;\;\;\;\;O^{2-}$
\newline .$\;\;\;\;\;\;\;\;\;\;\;\;\;\;\;\;\;\;\;\;\;\;\;\;\;\;\;\;\;\;\;\;\;Cu^{2+}\;\;O^{2-}\;\;\mathbf{Cu^{2+}}\;\;\mathbf{O^{2-}}\;\;\mathbf{Cu^{2+}}\;\;O^{2-}\;\;Cu^{2+}$
\newline .$\;\;\;\;\;\;\;\;\;\;\;\;\;\;\;\;\;\;\;\;\;\;\;\;\;\;\;\;\;
\;\;\;\;\;O^{2-}\;\;\;\;\;\;\;\;\;\;\;\;
\mathbf{O^{2-}}\;\;\;\;\;\mathring{O}\;\;\;\;\;\;\mathbf{O^{2-}}\;\;\;\;\;\;\;\;\;\;\;\;O^{2-}$
\newline .$\;\;\;\;\;\;\;\;\;\;\;\;\;\;\;\;\;\;\;\;\;\;\;\;\;\;\;\;\;\;\;\;\;\;Cu^{2+}\;\;O^{2-}\;\;\mathbf{Cu^{2+}}\;\;\mathbf{O^{2-}}\;\;\mathbf{Cu^{2+}}\;\;O^{2-}\;\;Cu^{2+}$
\newline .$\;\;\;\;\;\;\;\;\;\;\;\;\;\;\;\;\;\;\;\;\;\;\;\;\;\;\;\;\;\;
\;\;\;\;\;O^{2-}\;\;\;\;\;\;\;\;\;\;\;\;\;O^{2-}\;\;\;\;\;\;\;\;\;\;\;\;\;\;O^{2-}\;\;\;\;\;\;\;\;\;\;\;\;O^{2-}$
\bigskip 

\noindent The asymmetric positions of its four nearest neighbor 
$O^{2-}$ ions and four next-nearest neighbor 
$Cu^{2+}$ ions (all marked bold) give rise to EFG's at their nuclei and thus to NQR contributions. 
However, oppositely to hole-doped $La_{2-x}Sr_xCuO_4$, where the changes of NQR are caused by a change of hole density $n_p$ and $n_d$ in $O^{2-}$ and $Cu^{2+}$ orbitals at given EFG's, Eqs. (6, 5), the changes of NQR in the orbitals of each $\mathring{O}$ atom's neighbor ions, $O^{2-}$ and $Cu^{2+}$, are caused by \textit{changes of the EFG} at essentially constant $n_p \simeq n_{p0}$ and $n_d \simeq n_{d0}$ values. There are four such $O^{2-}$ and $Cu^{2+}$ neighbors to each $\mathring{O}$ atom with comparable distances and asymmetries. This necessitates a change of interpretation: In $HgBa_2CuO_{4+\delta}$ it is not a ``doped-hole'' density in the $O^{2-}$ and $Cu^{2+}$ neighbors of $\mathring{O}$, but the \textit{changed EFG} at their nuclei, that makes the finding  $P_p\simeq P_d\simeq 0.5$ plausible. The ``doped hole'' stays with the dopant $\mathring{O}$ atom, representing its open $2p^4$ shell, and contributes to NQR with its self-EFG.

In $YBa_2Cu_3O_{6+y}$, doped oxygen is incorporated (for $y>0.5$) in both the $CuO_2$ planes and the $CuO$ chains, Eq. (12). The NQR contributions from $\mathring{O}$ atoms and their neighbors in the $CuO_2$ planes are as already illustrated in the above diagram. An example for contributions from the $CuO$ chains, the diagram below gives an ion arrangement in the $Cu(1)$ plane as it may occur for ortho-III oxygen ordered ($y \approx 0.75$), with periodic repetition in the $a$ and $b$ direction. The nearest-neighbor configuration of the $O^{2-}$ ions and $O$ atoms is much more asymmetric than that of the $Cu^{2+}$ ions. Accordingly, much larger EFG's and NQR contributions can be expected from oxygen sites than from copper sites. Added to the essentially equal contributions from oxygen and copper sites in the $CuO_2$ planes, the finding of $P_p =0.64$ and $P_d =0.36$ for $YBa_2Cu_3O_{6+y}$ (Table III) becomes qualitatively plausible.

\bigskip 
\noindent .$\;\;\;\;\;\;\;\;\;\;\;\;\;\;\;\;\;\;\;\;\;\;\;\;\;\;\;\;\;\;\;\;\;\;\;\;\;\;\;\;\;\;\;Cu^{2+}\;\;\;\;\;\;\;\;\;\;\;Cu^{2+}\;\;\;\;\;\;\;\;\;\;\;Cu^{2+}$
\newline .$\;\;\;\;\;\;\;\;\;\;\;\;\;\;\;\;\;\;\;\;\;\;\;\;\;\;\;\;\;\;\;\;\;\;\;\;\;\;\;
\;\;\;\;\;O^{2-}\;\;\;\;\;\;\;\;\;\;\;\;\;\;O$
\newline .$\;\;\;\;\;\;\;\;\;\;\;\;\;\;\;\;\;\;\;\;\;\;\;\;\;\;\;\;\;\;\;\;\;\;\;\;\;\;\;\;\;\;\;Cu^{2+}\;\;\;\;\;\;\;\;\;\;\;Cu^{2+}\;\;\;\;\;\;\;\;\;\;\;Cu^{2+}$
\newline .$\;\;\;\;\;\;\;\;\;\;\;\;\;\;\;\;\;\;\;\;\;\;\;\;\;\;\;\;\;\;\;\;\;\;\;\;\;\;\;
\;\;\;\;\;O^{2-}\;\;\;\;\;\;\;\;\;\;\;\;\;\;O$
\newline .$\;\;\;\;\;\;\;\;\;\;\;\;\;\;\;\;\;\;\;\;\;\;\;\;\;\;\;\;\;\;\;\;\;\;\;\;\;\;\;\;\;\;\;Cu^{2+}\;\;\;\;\;\;\;\;\;\;\;Cu^{2+}\;\;\;\;\;\;\;\;\;\;\;Cu^{2+}$
\newline .$^{b}\uparrow\;\;\;\;\;\;\;\;\;\;\;\;\;\;\;\;\;\;\;\;\;\;\;\;\;\;\;\;\;\;\;\;\;\;\;
\;\;\;\;\;O^{2-}\;\;\;\;\;\;\;\;\;\;\;\;\;\;O$
\newline .$\;\;\;\;\rightarrow_a\;\;\;\;\;\;\;\;\;\;\;\;\;\;\;\;\;\;\;\;\;\;\;\;\;\;\;\;\;\;\;\;\;\;Cu^{2+}\;\;\;\;\;\;\;\;\;\;\;Cu^{2+}\;\;\;\;\;\;\;\;\;\;\;Cu^{2+}$

\bigskip 

\section{CONCLUSION}

NQR investigations and stripe analysis of high-$T_c$ cuprates complement each other: 
 (i) The small deviation of doped-hole density $h$ from the doping level of $La_{2-x}Sr_xCuO_4$, as observed with NQR, $\Delta{h} = x - h \approx 0.02$, agrees closely with the density of itinerant holes, $\tilde{p}$, responsible for suppression of 3D-AFM, as obtained from stripe incommensurability.
 (ii) The residence of doped holes at oxygen sites in $La_{2-x}Sr_xCuO_4$ and of doped electrons at copper sites in $Ln_{2-x}Ce_xCuO_4$ ($Ln = Pr, Nd$), as assumed in the stripe model, is (to a large degree) confirmed by the NQR studies. 
(iii) The NQR finding of doped-hole probabilities in oxygen and copper orbitals of oxygen-enriched high-$T_c$ cuprates, $P_p \simeq P_d \simeq \frac{1}{2}$, as well as of oxygen-doped $YBa_2Cu_3O_{6+y}$, $P_p \simeq 2P_d \simeq \frac{2}{3}$, is interpreted with the stripe model in terms of excess oxygen atoms in the $CuO_2$ planes and the $CuO$ chains.


\end{document}